\documentstyle[b98proc,psfig]{article}

\def\Journal#1#2#3#4{{#1} {\bf #2}, #3 (#4)}

\def\NPA{{\em Nucl. Phys.} A}

\def\PRL{\em Phys. Rev. Lett.}
\def\PREV{\em Phys. Rev.}

\def\ZPA{{\em Z. Phys.} A}

\newcommand{\be}{\begin{equation}}
\newcommand{\ee}{\end{equation}}

\def\ga{\mathrel{\mathpalette\fun >}}
\def\fun#1#2{\lower3.6pt\vbox{\baselineskip0pt\lineskip.9pt
\ialign{$\mathsurround=0pt#1\hfil##\hfil$\crcr#2\crcr\sim\crcr}}}

\begin{document}

\title{
PROPERTIES OF $\rho$-MESONS PRODUCED IN HEAVY ION COLLISIONS
}

\author{V. L. ELETSKY, B. L. IOFFE,}
\address{Institute for Theoretical and Experimental Physics, Moscow, Russia}
\author{J.I.KAPUSTA}
\address{University of Minnesota, Minneapolis, USA}

% You may repeat \author and \address if necessary

\maketitle

\abstracts{
The mass shift  and width broadening
 of $\rho$-mesons produced in heavy ion collisions is
estimated.
It is found that the mass increases by some tens of MeV but
the width becomes large, increasing by 200
MeV at beam energies of a few GeV$\cdot$A and by twice that
amount at beam energies of about a hundred GeV$\cdot$A.}

In a recent
paper \cite{ei}, two of us have argued that the mass
shift and width broadening of a particle in medium can be related to the
forward scattering amplitude $f(E)$ of this particle on the constituents of
the medium.
\be
\Delta m (E) = -2\pi\frac{\rho}{m}{\rm Re} f(E) \, , ~~~
\Delta \Gamma(E) = \frac{\rho}{m}~k\sigma(E) \, .
\label{dm}
\ee
Here $m$ is the vacuum mass of the particle, $E$ is its energy in the rest
frame of the constituent particle, $k$ is the particle momentum,
and $\rho$ is the density of consituents.
Eqs. (1) are applicable \cite{ei} if: \\
1) The particle's wavelength $\lambda$ is much less than the mean
distances between medium constituents $d$: $\lambda=k^{-1}\ll d$.
2) The particle's formation length $l_f$ is less than the nucleus radius $R$.
3) $\mid {\rm Re}f(E)\mid < d$. 4) The main part of
the scattering proceeds through small angles, $\theta \ll 1$. Only in this
case is the optical analogy on which Eqs. (1) are based correct.

We estimate the mass shift and width broadening in the
case of $\rho$-mesons produced in heavy ion collisions,
which can be observed
through the decay $\rho^0 \to e^+e^-$ or $\mu^+\mu^-$.  We will assume that
$\rho$-mesons are formed in the last stage of the evolution of hadronic
matter created in course of a heavy ion collision when the matter can be
considered as an almost noninteracting gas of pions and nucleons.
The main ingredients of our calculation
are $\rho\pi$ and $\rho N$ forward scattering amplitudes and total cross
sections as well as the values of nucleon and pion densities.
We consider central heavy ion collisions and assume that
nucleon and pion momentum distributions in the gas are just the momentum
distributions measured experimentally in such collisions.

To determine the amplitudes and cross sections
we use the following procedure.  At low energies we
saturate the cross sections and forward scattering amplitudes by resonance
contributions. At high energies we determine $\sigma_{\rho N}$ and
$\sigma_{\rho \pi}$ from $\sigma_{\gamma N}$  and $\sigma_{\gamma\pi}$
using the vector dominance model (VDM).  The cross section
$\sigma_{\gamma N}$  is well known experimentally,
${\rm Re}f_{\gamma N}$ is determined from the dispersion relation, and
$\sigma_{\gamma\pi}$ and ${\rm Re}f_{\gamma \pi}$ can be found by the Regge
approach.  Since VDM allows one to find only the cross sections
of transversally polarized $\rho$-mesons we restrict ourselves to this
case.  As was shown in \cite{ei}, when $E_{\rho}\ga 2$ GeV, $\Delta m$ and
$\Delta \Gamma$ for longitudinal $\rho$-mesons are much smaller than for
transversal ones in nuclear matter.  At zero $\rho$-meson energy,
$\Delta m$ and $\Delta \Gamma$  for transverse and longitudinal
$\rho$-mesons are evidently equal.
Therefore our results should be multiplied by a factor ranging
between 2/3 and 1 for unpolarized $\rho$-mesons.

To estimate ${\rm Re}_{\rho\pi}(E)$ and $\sigma_{\rho\pi}(E)$ at low
energy we take into account the following resonances:
 $a_1(1260)$, $\pi(1300)$, $a_2(1320)$ and $\omega(1420)$.
The nearest resonance under the threshold, $\omega(782)$, contributes
a negligilbe amount due to its narrow width.
According to Adler's theorem the pion scattering amplitude on any
hadronic target vanishes at zero pion energy in the target rest frame
in the limit of massless pions.
The corresponding factors were introduced in the resonance contributions.
At high energies we assume that the Regge approach is valid for $\gamma
\pi$  scattering and apply the vector dominance model (VDM) to relate $\rho
\pi$ and $\gamma \pi$ amplitudes. The Regge pole
contributions to $\sigma(s)$ and ${\rm Re}f(s)$
have the form:
\be
\sigma(s) = \sum_i ~r_i s^{\alpha_i-1} \, ,~~~
{\rm
Re}f(s) = -\frac{k}{4\pi s}\sum_i~\frac{1+\cos \pi \alpha_i}{\sin \pi
\alpha_i}r_is^{\alpha_i} \, .
\ee
For $\sigma_{\gamma \pi}$  only $P$ (Pomeron) and $P^{\prime}$ Regge poles
contribute \cite{bo,bk}.
The residues of the $P$ and $P^{\prime}$ poles in
$\gamma\pi$  scattering were found in
\cite{bk} using Regge pole factorization and data on $\gamma p$, $\pi p$
and $pp$ scattering.
\be
\sigma_{\pi \gamma}(s) = 7.48 \alpha \left [
(s/s_0)^{\alpha_P-1} + 0.971 (s/s_0)^{\alpha_{P^{\prime}}-1} \right ] \, ,
\ee
where
$\alpha_P=1.0808,~\alpha_{P^{\prime}}=0.5475$, $\alpha=1/137$, $s_0=1$
GeV$^2$ and $\sigma$ in Eq. (5) is given in millibarns.

For the amplitude ${\rm Re} f_{\rho N}$ at laboratory energies of the $\rho$
above 2 GeV we use the results obtained \cite{ei} with the
dispersion relation, VDM and experimental data on $\sigma_{\gamma N}$.  At
lower energies we again use the resonance approximation and take 10 $N$ and
$\Delta$ resonances with significant branchings into $\rho N$ and with masses
above the $\rho N$ threshold  and below 2200 MeV.  Besides these resonances,
two others with masses below the $\rho N$ threshold were accounted for: the
$\Delta(1238)$ and the $N(1500)$.  It was assumed that VDM is valid for the
contribution of these resonances to the widths $\Gamma_{\rho N}$ and
$\Gamma_{\gamma N}$.

The results for $\sigma_{\rho\pi}$, ${\rm Re}f_{\rho\pi}$,
$\sigma_{\rho N}$ and ${\rm Re} f_{\rho N}$ in the rest frame of the target,
are shown in Figs.1 and 2.

\newpage

\begin{figure}[ht]
\begin{minipage}{0.47\textwidth}
{{\hspace{3mm}a)\hspace{-3mm}}
\psfig{file=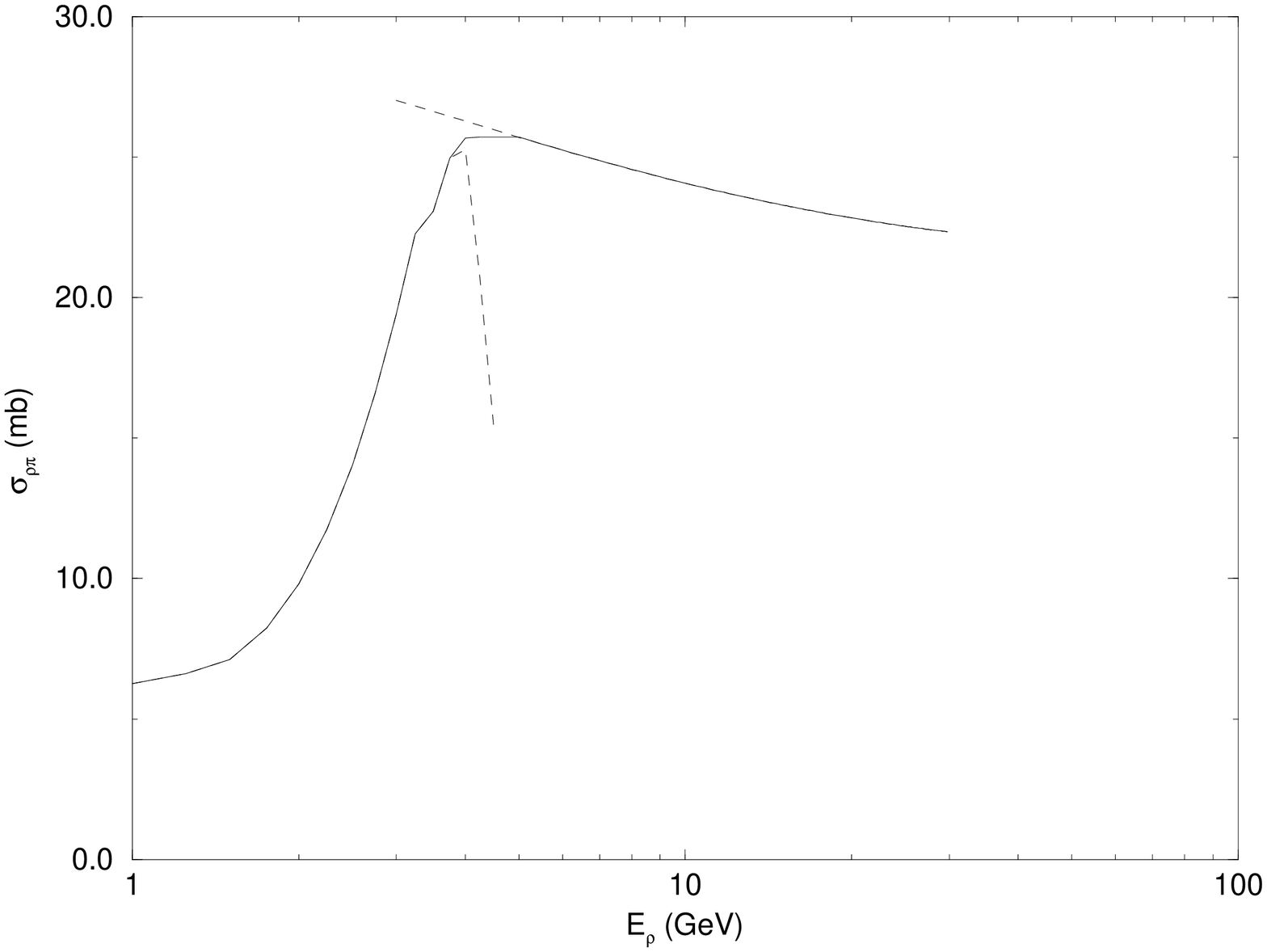,width=1.0\textwidth,angle=-90}}
\end{minipage}
\begin{minipage}{0.47\textwidth}
{{\hspace{3mm}b)\hspace{-3mm}}
\psfig{file=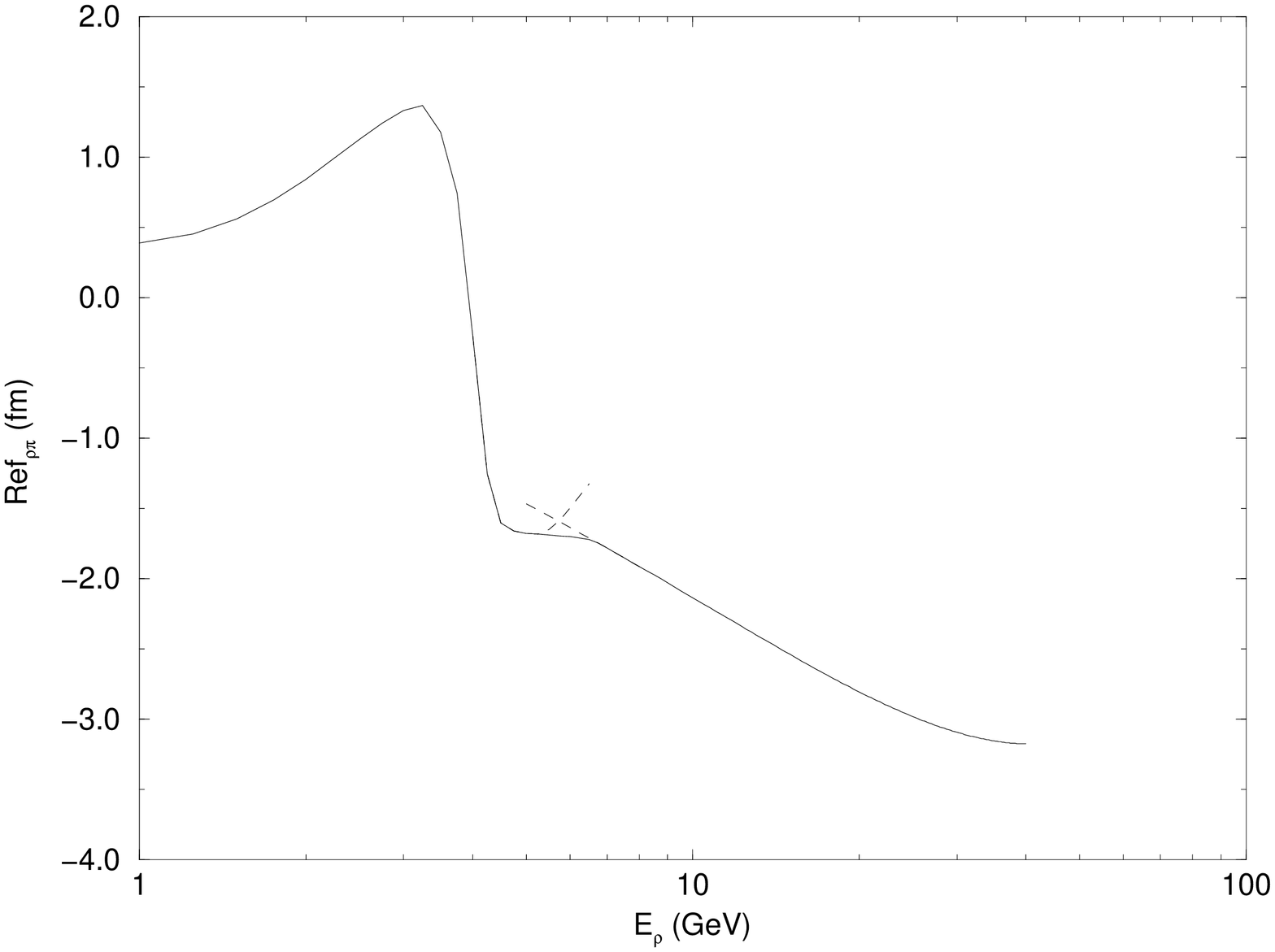,width=1.0\textwidth,angle=-90}}
\end{minipage}
\caption{Cross section (a) and real part of the
forward scattering amplitude (b) for $\rho$-mesons
scattering on pions as functions of the total $\rho$-meson energy
in the pion rest frame.  The curves at low energy are the result of
the resonance approximation.  The curves at high energy are the
result of the Regge parametrization.  These curves are matched at
intermediate energies.}
\end{figure}

\begin{figure}[ht]
\begin{minipage}{0.47\textwidth}
{{\hspace{3mm}a)\hspace{-3mm}}
\psfig{file=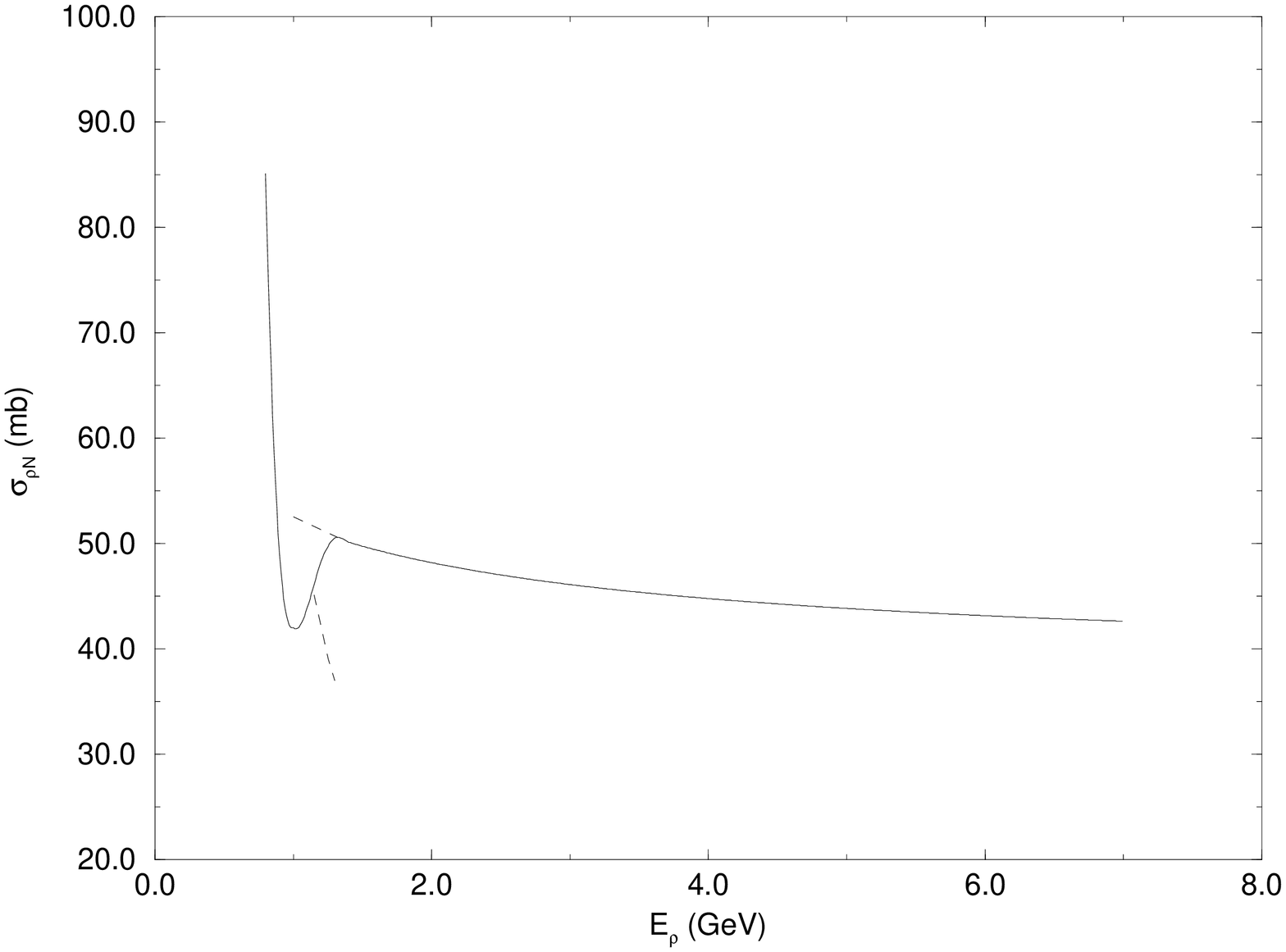,width=1.0\textwidth,angle=-90}}
\end{minipage}
\begin{minipage}{0.47\textwidth}
{{\hspace{3mm}b)\hspace{-3mm}}
\psfig{file=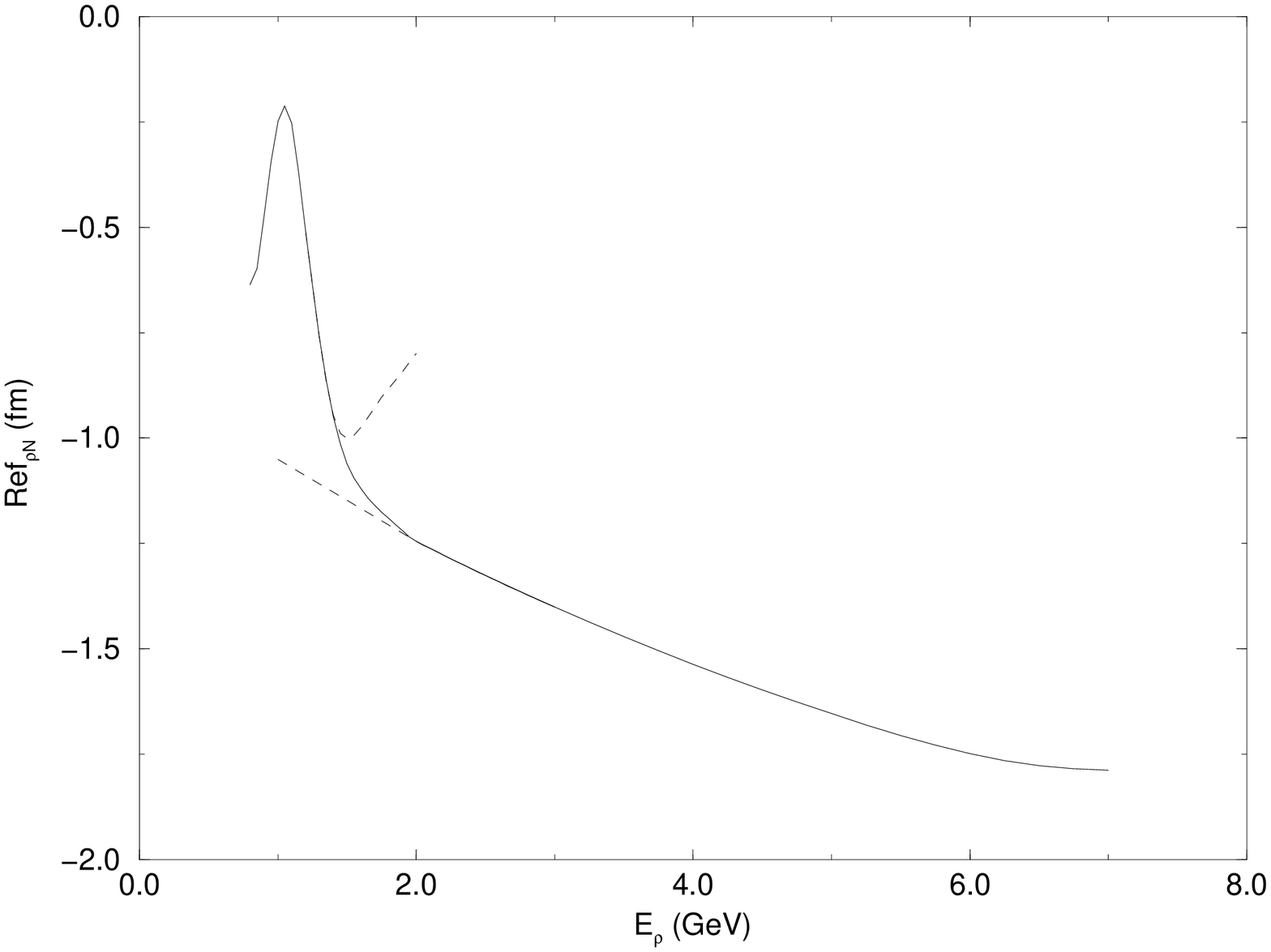,width=1.0\textwidth,angle=-90}}
\end{minipage}
\caption{Same as Fig.1, but for $\rho$-mesons scattering on
nucleons. The curves at high energy are from Ref.$^1$
%\cite{bo}
.}
\end{figure}

As shown by the experimental data, the nucleons and pions produced in
heavy ion collisions cannot be considered as a gas in global thermal
equilibrium even during the last
stage of evolution of hadronic matter created in the collisions.
In the case of
high energy collisions the longitudinal and
transverse momenta of nucleons and pions are very different. In the
experiment on $S + S$  collisions at 200 GeV$\cdot$A \cite{ba} it was
found that $\langle p^{\rm cm}_{LN}\rangle =3.3$ GeV,
$\langle p_{TN}\rangle = 0.61$ GeV, and $\langle p^{\rm cm}_{L\pi}\rangle
\approx 0.70$ GeV, $\langle p_{T\pi}\rangle
\approx 0.36$ GeV.

The  angular distributions of pions produced in $Ni+Ni$  collisions at
$E=1-2$ GeV$\cdot$A shows essential anisotropy \cite{pe}. If the pion
angular distribution in the centre of mass system is approximated by $1+
a\cos^2 \theta$ then, from the data, follows $a \approx 1.3$.

When calculating the $\rho$-meson mass shift and
width broadening an averaging must be performed over the $\rho$-meson
direction of flight relative to nucleons and pions. Such a calculation can
be done only for real experimental conditions. For this reason we restrict
ourselves to rough estimates.
As an example take the
experiment \cite{ba} for central collisions where the ratio of
pion to nucleon multiplicities was found to be $R_{\pi}=5.3$.  Suppose
that in this experiment the $\rho$-meson is produced with longitudinal and
transverse momenta in the laboratory system $k_{L}=3$ GeV,
$k_{T}=0.5$ GeV.
For these values of $\rho$ momenta the formation time of the $\rho$-meson
is close to the mean formation time of pions so a necessary condition
of our approach is fulfilled.
Then it is possible, using the curves of Figs. 1 and 2
to calculate the mean values:
$\langle {\rm Re}f_{\rho N} \rangle \approx -1.1~{\rm fm}$,
$\langle {\rm Re}f_{\rho\pi} \rangle \approx 0.03~{\rm fm}$,
$\langle \sigma_{\rho N} \rangle \approx 45 ~{\rm mb}$,
$\langle \sigma_{\rho\pi} \rangle \approx 20~{\rm mb}$.
The small value of $\langle{\rm Re}f_{\rho\pi}\rangle$ arises from a
compensation of positive and negative contributions from low and
high energy collisions.
For the nucleon and pion densities we take
\be
\rho_N= \rho^0_N/(1 + R_{\pi}\beta)\, , ~~~\rho_{\pi} =
\rho^0_N R_{\pi}/(1+R_{\pi}\beta)\, ,
\ee
where $\rho^0_N=1/\upsilon_N$ and $\beta=\upsilon_{\pi}/\upsilon_N$
and it is assumed that at this
stage of evolution any participant -- nucleon or
pion -- occupies the fixed volume $\upsilon_N$ or $\upsilon_{\pi}$,
respectively.
For numerical estimates we take
$\rho^0_N=0.3$ fm$^{-3}$, about two times standard nucleon density.
Using (1) and (4) together with the experimental values $R_{\pi} = 5.3$
and $\beta=1$, we get
\be
\Delta m_{\rho} = 18 - 2 = 16~{\rm MeV},~~~~
\Delta \Gamma_{\rho}\approx 150 + 400 = 550~{\rm MeV}
\ee
The first numbers above refer to the contributions from $\rho-N$ and second
from $\rho-\pi$  scattering. Because the $\rho$-meson width broadening
appears to be very large, a basic condition of our approach, $\Delta \Gamma
\ll m_{\rho}$, is badly fulfilled. The applicability condition of the
method, $\mid{\rm Re}f\mid < d$, is not well satisfied either since in this
case $d=0.9$ fm. For these reasons the values of $\Delta m_{\rho}$ and
$\Delta\Gamma_{\rho}$ may be considered only as estimates.

The main uncertainty in our approach comes from
the assumed value of the nucleon density at the final stage of
evolution: $\rho^0_N =0.3$ fm$^{-3}$. If this density would
be a factor of two smaller then $\Delta\Gamma_{\rho}\sim 250$ MeV and
the $\rho$-meson could be observed as a broad
peak in the $e^+e^-$  or $\mu^+\mu^-$ mass spectrum.

Our qualitative conclusion is that in central collisions of heavy nuclei at
high energies, $E\sim 100$ GeV$\cdot$A, where a large number of pions per
participating nucleon is produced, the $\rho$-peak will be
observed in $e^+e^-$ or $\mu^+\mu^-$ mass distributions only as a
very broad enhancement, or even no enhancement at all.  Inspite of
the assumptions we made, including noninteracting nucleon and pion
matter at the final stage of evolution and the specific numerical
value of the nucleon density, we believe that this qualitative
conclusion is still valid.  This conclusion
is in qualitative agreement with the measurement of $e^+e^-$  pair
production in heavy ion collisions \cite{ag} where no $\rho$-peak was found and
only a smooth $e^+e^-$ mass spectrum from 0 to 1 GeV was observed.
If, however, such a peak would be observed in future experiments
it would indicate that the hadronic (nucleon and pion) density at the
final stage of evolution, where the $\rho$-meson is formed, is very low,
even lower than normal nuclear density.

Let us turn now to the case of lower energy heavy ion collisions,
$E \sim$ a few GeV$\cdot$A. Consider, as an example, heavy ion collisions at
$E_{\rm kin}=3$ GeV$\cdot$A and production of $\rho$-mesons of energy
$E^{\rm tot}_{\rho}=1.2$ GeV in the forward direction.
The number of pions produced can be estimated by extrapolation of the data
\cite{pe} on $Ni+Ni$  collisions.
We find that $R_{\pi}=0.48$.
At
such low energies it is reasonable to suppose that for pions
$\langle p_L \rangle= \langle p_{\perp}\rangle
\approx 0.2$ GeV and for nucleons $\langle p_{TN}\rangle=0.61$ GeV.
Then we obtain the mean values:
$\langle {\rm Re}f_{\rho N} \rangle= -0.54~{\rm fm}$,
$\langle {\rm Re}f_{\rho \pi} \rangle= 0.30~{\rm fm}$,
$\langle \sigma_{\rho N}\rangle = 45~{\rm mb}$,
$\langle \sigma_{\rho \pi} \rangle= 13~{\rm mb}$.
For the $\rho$-meson mass shift and width broadening we have:
\be
\Delta m_{\rho} = 37 - 10 = 27~{\rm MeV},~~~~
\Delta\Gamma_{\rho} = 245 + 35 = 280~{\rm MeV}.
\ee
The first numbers above refer to $\rho N$ scattering, the second
ones to $\rho\pi$. The conclusion is that in low energy heavy ion
collisions a $\rho$-peak may be observed in $e^+e^-$ or $\mu^+\mu^-$
mass distributions as a broad enhancement approximately at the position of
$\rho$-mass.

This work was supported by the CRDF grant RP2-132,
Schweizerischer National Fonds grant 7SUPJ048716, RFBR grant
97-02-16131, and U.S. Department of Energy grant DE-FG02-87ER40328.
V.E. acknowledges support of BMBF, Bonn, Germany. B.I. is thankful to
A.v.Humboldt Foundation for financial support which allowed him to
participate in the Conference.

%\section*{Figure}

%\section{A section about figures}

%Use EPS and a figure environment for pictures like fig.~\ref{bar}.
%\begin{figure}[ht]
%\psfig{figure=figure.eps,height=2cm,width=15cm}
%{a) Cross section (left are,upper curve) and real part of the
%forward scattering amplitude (right exe,lower curve) for $\rho$-mesons
%scattering on pions as functions of the total $\rho$-meson energy in the
%pion rest frame.  The curves at low energy are the result of the resonance
%approximation.  The curves at high energy are the result of the Regge
%parametrization.  These curves are matched at intermediate energies.
%b) Same as a) but for $\rho$-mesons scattering on
%nucleons.  The curves at high energy are from Ref.\cite{bo}.}
%\label{bar}
%\end{figure}

%\subsection{A Subsection}
%Just to show a subsection.

%\section{Summary from the abstract booklet}

% Please use the \Journal macro together with \PLA,\PLB etc. whenever possible!
%\bibitem{EXAMPLE1}A. N. Onymous and U. N. Known,
%\Journal{\PLB}{11}{1111}{1998} \end{thebibliography}

\end{document}